\documentstyle[multicol,aps]{revtex}

\begin{document}

\title{Composite Fermions with Orbital Magnetization}

\author{Steven H.  Simon${}^1$, Ady Stern${}^2$, and Bertrand I.
  Halperin${}^3$} \address{${}^1$Department of Physics, Massachusetts
  Institute of Technology, Cambridge, MA 02139 \\ ${}^2$Department of
  Condensed Matter Physics, Weizmann Institute of Sciences, Rehovot
  76100, Israel \\ ${}^3$Department of Physics, Harvard University,
  Cambridge, MA 02138}


\maketitle

\begin{abstract}
  For quantum Hall systems, in the limit of large magnetic field $B$
  (or equivalently small electron band mass $m_b$), the static
  response of electrons to a spatially varying magnetic field is
  largely determined by kinetic energy considerations.  This response
  is not correctly given in existing approximations based on the
  Fermion Chern-Simons theory of the partially filled Landau level.
  We remedy this problem by attaching an orbital magnetization to each
  fermion to separate the current into magnetization and transport
  contributions, associated with the cyclotron and guiding center
  motions respectively.  This leads to a Chern-Simons Fermi liquid
  description of the $\nu=\frac{1}{2m}$ state which correctly predicts
  the $m_b$ dependence of the static and dynamic response in the limit
  $m_b \rightarrow 0$.
\end{abstract}



\begin{multicols}{2}

A useful theoretical tool for studying quantum Hall states is the
composite fermions picture, known also as the Chern--Simons fermionic
theory\cite{HLR}.  Based on the work of Jain\cite{Jain}, the
Chern-Simons fermion picture was introduced by Lopez and
Fradkin\cite{Lopez} to study fractional quantized Hall states, and by
Halperin, Lee and Read (HLR)\cite{HLR}, as well as Kalmeyer and
Zhang\cite{Zhang}, to study even denominator filling fractions.

In the composite fermion description, a system of spin polarized
electrons in a two dimensional electron gas is transformed exactly
into a system of composite, or gauge-transformed, fermions.  These
composite fermions carry an even number $\tilde \phi = 2m$ of
`Chern-Simons' flux quanta and interact with each other both
electrostatically and via the Chern--Simons gauge field\cite{HLR}.  At
a mean field level the fermions are subject to a uniform effective
magnetic field $ \Delta B= B-\Phi_0 {n_e}\tilde{\phi} $ where
$\Phi_0\equiv h/e$ is the flux quantum, and $n_e$ is the average
electron density (the speed of light $c=1$).  Such an approximation
maps electrons at filling fraction $\nu=\frac{1}{2m}$ to fermions in
zero magnetic field, where the ground state should be a gapless Fermi
liquid-like state.  Similarly, this transformation
maps\cite{HLR,Lopez} fractional quantized Hall states at filling
fractions $\nu = \frac{p}{2mp+1}$ to integer quantized Hall states of
fermions at $\nu = p$.

The main systematic attempts for going beyond mean field theory have
so far involved perturbative treatments of the Chern--Simons and
electrostatic interactions.\cite{HLR,Simonhalp,Ady,Kim} Attention has
focused on one difficulty --- the divergence of the composite
fermion's effective mass at the Fermi surface when the range of
electrostatic interaction is shorter than, or equal to, that of the
Coulomb interaction.  This divergence, due to infra-red gauge field
fluctuations, is manifested in the energy gaps of fractional quantized
Hall states at $\nu=\frac{p}{2mp+1}$ (for large $p$) but does not
affect the electronic linear response at $\nu=\frac{1}{2m}$ at zero
temperature, due to a mutual cancelation with another singular
term\cite{Ady,Kim}.  Consequently\cite{Ady} the low energy excitations
at $\nu = \frac{1}{2m}$ are best characterized by another, {\it
  finite}, effective mass, denoted by $m^*$, which is the effective
mass of relevance to the present work.  It is this $m^*$ which should
determine the scale of the fractional Hall gaps for small values of
$p$.

In this paper we consider a second source of difficulty in the
Chern-Simons theory which appears even when gauge-field fluctuations
are not infra-red singular.  (For example, there is no predicted
infra-red divergences in the effective mass if the electron-electron
repulsion falls off more slowly than $1/r$.)  This second difficulty
is encountered in the limit of a small band mass $m_b$ or equivalently
the large $B$ limit.  The fact that, in this limit, the electronic
ground state and low energy excitations are constrained to the lowest
Landau level, leads to certain features of the electronic response to
an external static vector potential which are not properly embodied in
previous approximation schemes.  In this paper we suggest a remedy to
that flaw.  We argue that while previously used approximations are
appropriate for calculating the density-density response function
$K_{00}$ (defined below) they are not appropriate for response
functions such as the density response to a transverse vector
potential $K_{01}$. The approach we present allows us to derive an
approximation to the linear response function which yields the correct
$m_b$ dependence in the limit $m_b\rightarrow 0$.

Our approach is based on a separation of the current into a
magnetization current which is associated with the cyclotron motion of
electrons and a transport current associated with the guiding center
motion.  This separation is achieved by attaching a magnetization
${\mu_{\mbox{\tiny M}}}$ to each particle.  This magnetization
originates from the electrons' orbital motion and is unrelated to the
spin (which is assumed to be a frozen degree of freedom).  In the
limit $m_b \rightarrow 0$, the magnetization ${\mu_{\mbox{\tiny M}}}$
is given by the Bohr magneton $ \mu_b = e \hbar/ (2 m_b)$.  Our
separation procedure, combined with approximations similar to those
made in previous works, results in response functions that correctly
describe the $m_b \rightarrow 0$ limit.

We will mostly discuss the behavior of the electromagnetic response
matrix $K_{\mu\nu}$ which is closely related to the
conductivity\cite{HLR,Simonhalp}.  To define $K$, a weak vector
potential $A_\mu^{\mbox{\scriptsize{ext}}}$ is externally applied to a
system at wavevector ${\bf q}$ and frequency $\omega$, and
consequently, a current $j_\mu$ is induced (Here $A_0$ is the scalar
potential, and $j_0$ is the induced density).  We write the response
function in the form $j_{\mu}(q, \omega) = K_{\mu \nu}(q, \omega)
A_{\nu}^{\mbox{\scriptsize{ext}}}(q, \omega)$ where $\mu$ and $\nu$
take the values $0,x,y$.  We will use the convention that the
perturbation is applied with $q \| {\bf {\hat x}}$ so that the
longitudinal current is $j_x = (\omega/q) j_0$.  Using the gauge ${\bf
  A}_x = 0$, we can then treat $K_{\mu\nu}$ as a $2\times 2$ matrix
with indices taking the values 0 or 1 denoting the time or transverse
space components.  In this notation the current vector $j_\mu$ is
$(j_0,j_y)$, and the vector potential $A_\mu$ is $(A_0,A_y)$.  In the
HLR approach\cite{HLR}, the response matrix $K$ is calculated within
the RPA.  As reviewed below, an improved `Modified' RPA (MRPA) was
introduced in later work\cite{Simonhalp}.

To understand why we believe the RPA and MRPA miss some important
physics, we consider a system at $\nu = \frac{1}{2}$ such that the
magnetic field is $B_{1/2} = 2 n_e \Phi_0$. We now apply a weak static
field $\delta B({\bf r})$ at some small but finite wavevector $q$.  At
a mean field, RPA, or MRPA level, the transformed fermions see only the
additional field $\delta B({\bf  r})$, and the energy is the same in
regions where $\delta B > 0$ as it is in regions where $\delta B < 0$.
However, the original electrons see the full field $B_{1/2} + \delta
B({\bf  r})$.  Since the electrons are in the lowest Landau level they
have a kinetic energy $\frac{1}{2} \hbar \omega_c({\bf  r}) = \mu_b
(B_{1/2} +\delta B({\bf  r}))$ which is lower in regions where $\delta B
< 0$.  In the limit of $m_b \rightarrow 0$, the electrons are then
strongly attracted to regions where $\delta B({\bf  r}) < 0$.  This
behavior should be reflected in $K_{01}$, but is not accounted for at
a mean field or (M)RPA level.

In order to model this attraction to regions of minimal magnetic
field, we attach to each particle a fixed magnetization
${\mu_{\mbox{\tiny M}}}$ which interacts with the magnetic field and
creates an energy cost ${\mu_{\mbox{\tiny M}}} B({\bf r})$ as a
function of position.  In the limit where the interaction energy goes
to zero (or equivalently if $m_b \rightarrow 0$), ${\mu_{\mbox{\tiny
      M}}}$ is given by $\mu_b$, so that this energy cost consists of
the kinetic energy only and is given by ${\mu_{\mbox{\tiny M}}} B({\bf
  r}) = \frac{1}{2} \hbar \omega_c({\bf r})$.  For finite interaction
energy the local energy cost is composed of kinetic and interaction
contributions.  Within our framework, most of that energy should be
described as magnetization energy, with the remaining part described
as quasiparticle interaction energy.  There is, however, some
arbitrariness in that separation, and consequently in the definition
of ${\mu_{\mbox{\tiny M}}}$.  For the present purpose we need to
consider only the leading behavior in $1/m_b$ and we can therefore put
${\mu_{\mbox{\tiny M}}} = \mu_b$.  Following the attachment of
magnetization, we define an effective scalar potential seen by the
magnetized particles as $ A_0^{\mbox{\scriptsize{eff}}} = A_0 +
{\mu_{\mbox{\tiny M}}} B({\bf r})$ and declare that the particles
respond to this effective field.

Another effect that is not accounted for in previous works involves
the magnetization current, whose physical interpretation is the
following.  In a high magnetic field, each particle can be thought of
as traveling in a cyclotron orbit.  When the density of particles is
uniform, the local currents of these orbits cancel and there is no net
current in the bulk of the system.  However, when there is a density
inhomogeneity, these local currents do not quite cancel and a net
magnetization current results.  Formally, we write $ {\bf
  j}_{\mbox{\scriptsize{mag}}} = \nabla \times {\bf M}$ where ${\bf
  M}$ is the magnetization density which is given, for noninteracting
particles in the lowest Landau level, by ${\bf M} = \frac{1}{2}
{\bf{\hat z}} \hbar \omega_c({\bf r}) n({\bf r}) / B({\bf r}) = \mu_b
n({\bf r})$ (where $n({\bf r}) = n_e + j_0({\bf r})$ is the local
density).  Again, more generally $\mu_b$ might be replaced by
${\mu_{\mbox{\tiny M}}}$ such that ${\bf M} = {\mu_{\mbox{\tiny M}}}
n({\bf r})$.  Thus, the magnetization current is $ {\bf
  j}_{\mbox{\scriptsize{mag}}} = {\mu_{\mbox{\tiny M}}} ({\bf {\hat z}}
\times \nabla n({\bf r}))$.  By attaching magnetization
${\mu_{\mbox{\tiny M}}}$ to each particle, which is equivalent to
attaching a current loop, we separate out the contribution of the
magnetization current to the total current.  It is then convenient to
define the transport\cite{Thermo} current ${\bf
  j}_{\mbox{\scriptsize{trans}}} = {\bf j}_{\mbox{\scriptsize{total}}}
- {\bf j}_{\mbox{\scriptsize{mag}}} $.  The transport current may be
interpreted as the current of the bound particle-magnetization
composite objects.

When projected to the lowest Landau level, the projected current and
density operators satisfy $P {\bf j} P = \mu_b ({\bf {\hat z}} \times
\nabla P n P)$, ie, the projected current operator describes only
magnetization current\cite{LongerPaper} (Here, $P$ is the projection
operator).  Therefore, in the limit $m_b \rightarrow 0$, if we apply a
static scalar potential $A_{0}^{\mbox{\scriptsize{ext}}}(q)$ to the
system and we look at the current response in powers of $1/m_b$, we
find a transverse magnetization current $\mu_b {\bf{\hat z}} \times i
{\bf q} K_{00} A_{0}^{\mbox{\scriptsize{ext}}}$ originating from
lowest Landau level matrix elements, as well as a current independent
of $m_b$, which may be described as arising from virtual transitions
to other Landau levels.  Thus, if $q$ is finite and $m_b \rightarrow
0$, we expect $K_{10}/K_{00} \rightarrow i q \mu_b$.  This result is
not contained in previous works based on the Chern-Simons approach.

Before detailing our remedies to the above problems, we review the RPA
and MRPA.  Both approaches separate the long ranged interactions by
writing $K$  as\cite{Simonhalp,Ady,LongerPaper,Pines}
\begin{equation}
  \label{eq:KPi}
  K^{-1} = \Pi^{-1} + U ~~~~ \mbox{where}~~~~U = \left(
  \begin{array}{ccc} v(q) & -c(q)
\\  c(q)  & 0 \end{array} \right)_.
\end{equation}
Here $v(q) = e^2/(\epsilon q)$ is the Coulomb interaction and $c(q) =
i \tilde \phi \Phi_0/q$ is the Chern-Simons interaction.  In other
words, $\Pi$ is the part of $K$ that is irreducible with respect to
both $v(q)$ and $c(q)$.  The RPA\cite{HLR,Lopez} is obtained by
setting $\Pi$ equal to $K^0$, the response of a noninteracting system
of fermions with mass $m_b$ in the mean magnetic field $\Delta B$.
This RPA approach has the problem that the low energy excitations are
on a scale set by the cyclotron energy (ie, by $m_b$) rather than by
the interaction energy.  The MRPA\cite{Simonhalp} repairs this problem
by defining an effective mass $m^*$ that is set phenomenologically by
the interaction scale\cite{HLR}.  Galilean invariance is maintained by
adding a Landau interaction term ${\cal F}_1$ to insure that Kohn's
theorem and the $f$-sum rule will be satisfied.  To define the MRPA we
write
\begin{eqnarray}
  \label{eq:Pi*}
  \Pi^{-1} &=& [\Pi^*]^{-1} + {\cal F}_1 \\
  \label{eq:F1def} {\cal F}_1 &=&
  {\frac{(m^*\!-\!m_b)}{n_e e^2}} \left(
  \begin{array}{cc}  {\frac{\omega^2}{q^2}} & 0 \\ 0 & -1 \end{array}
  \right)_.
\end{eqnarray}
The MRPA is then obtained by setting $\Pi^*$ equal to the response
$K^{0*}$ of a system of noninteracting fermions of mass $m^*$ in the
mean magnetic field $\Delta B$.  The response function thus calculated
(using $\Pi^* =K^{0*}$ and Eqns.~\ref{eq:Pi*} and \ref{eq:KPi}) will
be called $K^{{\mbox{\tiny{MRPA}}}}$.  Comparisons of numerical
results of exact diagonalizations to results of $K_{00}$ calculated in
the MRPA were quite favorable\cite{Simonhalp} for the low energy
excitations at $\nu=\frac{p}{2mp+1}$ for small $p$.

As described above, the (M)RPA approach does not account for
magnetization effects due to the fact that when we take the mean field
solution as a starting point for a perturbation theory for the
Chern-Simons fermions, we lose the fact that the original electrons
travel in local cyclotron orbits.  This piece of physics would
presumably be recovered if one could correctly carry out a
renormalization procedure which eliminates the high frequency motion
at the cyclotron frequency to obtain equations valid on the scale of
the Coulomb interaction.  In the current work, we recover this physics
by attaching magnetization to each particle by hand.  This attachment
is not an exact transformation, but is rather a way of modeling
behavior that is lost when we take the mean field as a starting point.
As we will see below, within a Landau-Fermi liquid theory picture,
this attachment seems to give the correct quasiparticles for the
system.

The attachment of magnetization to the particles does not change the
interparticle interactions.  However, the magnetized fermions now
respond to the effective potential and the motion of these magnetized
fermions yields only the transport current response.  We thus define a
matrix $\tilde K$ to be the {\it transport} current response of the
electrons to the external {\it effective} potential.  We thus write
\begin{equation}
     \begin{array}{cc} A_{\mbox{\scriptsize{eff}}} =&
       M^\dagger A \\ j_{\mbox{\scriptsize{total}}} =& M
       j_{\mbox{\scriptsize{trans}}}
       \end{array}  ~~~~~~~~
    \label{eq:Mdef2} M = \left(
    \begin{array}{cc} 1 ~& 0 \\ i q {\mu_{\mbox{\tiny M}}} & 1
\end{array} \right)
\end{equation}
and relate $\tilde K$ to $K$ by $ K = M \tilde K M^\dagger$.  As
discussed above, in the limit $m_b \rightarrow 0$, we must have
${\mu_{\mbox{\tiny M}}} \rightarrow \mu_b$.  In the rest of this
paper, however, we will consider ${\mu_{\mbox{\tiny M}}} = \mu_b$.

We now propose an approximation which we call the `Magnetized Modified
RPA' or $\mbox{M}^2$RPA where we set $\tilde K$ equal to
$K^{{\mbox{\tiny{MRPA}}}}$.  Thus in the $\mbox{M}^2$RPA we
approximate
\begin{equation}
  K \approx M \left( [K^0{}^*]^{-1} + {\cal F}_1 + U \right)^{-1}
  M^\dagger.
\end{equation}
This equation is a central result of this paper.  In the limit $m_b
\rightarrow 0$, the $\mbox{M}^2$RPA correctly describes the static
response properties described above.  It should be noted that the
$\mbox{M}^2$RPA predicts $K_{00} = K_{00}^{{\mbox{\tiny{MRPA}}}}$ and
is therefore equally well supported by exact
diagonalizations\cite{Simonhalp}.  Furthermore, we note that the
excitation modes of a system are determined by the poles of
$\mbox{det}[K]$, and since $\mbox{det}[K] = \mbox{det}[\tilde K]$, the
MRPA and the $\mbox{M}^2$RPA predict the same excitation energies.
However, the $\mbox{M}^2$RPA calculations of $K_{01}$ and $K_{10}$
differ from $K_{01}^{{\mbox{\tiny{MRPA}}}}$ and
$K_{10}^{{\mbox{\tiny{MRPA}}}}$ by terms of order $q/m_b$ and the
$\mbox{M}^2$RPA value of $K_{11}$ differs from
$K_{11}^{{\mbox{\tiny{MRPA}}}}$ by terms of order $q/m_b$ and by terms
of order $(q/m_b)^2$.  It should be noted that all finite $q$
experimental tests\cite{WillettReview} of the Chern-Simons theory to
date have measured only $K_{00}$.  Similar to MRPA, we expect the
$\mbox{M}^2$RPA, in addition to describing the $\nu=\frac{1}{2m}$
Fermi liquid states, should properly describe the Jain series of
quantized states $\nu=\frac{p}{2mp+1}$ for small $p$.  At large values
of $p$, the description should be modified to account for the effects
of the singular infra-red gauge fluctuations.

We now turn to discuss how the $\mbox{M}^2$RPA fits into the general
picture of a Fermi liquid theory of the $\nu=1/(2m)$ state.  In
essence, we show that $\mbox{M}^2$RPA amounts to adopting the Fermi
liquid picture of Ref.~\onlinecite{Ady} as describing the dynamics of
magnetized composite fermion quasiparticles rather than unmagnetized
ones.

In Fermi liquid theory for fermions with short ranged interactions,
such as ${}^3$He, the response function $K$ is described by the
solution of a Landau-Boltzmann equation\cite{Pines}.  However, for
fermions with long ranged interactions\cite{Pines}, the Silin
extension of the Landau theory asserts that it is not the function $K$
that is described by the Landau-Boltzmann equation, but rather a
response function called $\Pi$.  In the case of electrons at zero
magnetic field in 2D, for example, $\Pi$ is defined by
Eq.~\ref{eq:KPi}, but with $c(q) = 0$.  Eq.~\ref{eq:KPi} separates out
the Hartree part of the diverging long ranged interaction such that
$\Pi$ gives the quasiparticle response to the sum of the external
vector potential and the induced internal vector potential.  Within
this theory, the Landau-Boltzmann equation describes the dynamics of
quasiparticles of mass $m^*$ near the Fermi
surface\cite{Simonhalp,Ady,LongerPaper,Pines}.  The short ranged
effective interactions between two quasiparticles whose momenta differ
by an angle $\theta$ is described by a function $f(\theta)$. The
`Landau coefficients' are defined as $f_l = \frac{1}{2\pi} \int_{0}^{2
  \pi} d \theta f(\theta) e^{i l \theta}$ (by symmetry, $f_l =
f_{-l}$).  The parameters $f_0$ and $f_1$ are fixed by the
identities\cite{Pines} \vspace{-.1in}
\begin{equation}
  \frac{d \mu}{d n} = \frac{2 \pi \hbar^2}{m^*} + f_0 ~~~~~~ ;
  ~~~~~\frac{1}{m_b} = \frac{1}{m^*} + \frac{f_1}{2 \pi \hbar^2}_.
  \label{eq:frules}
\end{equation}
The second identity is a result of Galilean invariance\cite{Pines}.

Since in the Chern-Simons problem, the interactions are also long
ranged, we have similarly made the separation\cite{HLR} defined in
Eq.~\ref{eq:KPi} (with $c(q) = i \Phi_0 \tilde \phi/q$).  However, in
the limit $m_b \rightarrow 0$, further separation should be carried
out to remove the diverging magnetization effects.  To this end we
define a response function $\tilde \Pi$ by
\begin{equation}
  \label{eq:tildePi}
  \Pi = M \tilde \Pi M^\dagger.
\end{equation}
definition, $\tilde \Pi$ relates the transport current of the {\it
  magnetized} quasiparticles to the effective vector potential
(including both external and internally induced
contributions\cite{LongerPaper}).  It is $\tilde \Pi$ which we claim
is given by a Landau-Boltzmann equation describing the dynamics of
quasiparticles with the finite effective mass $m^*$.

Recent works attempting to formulate a Landau-Boltzmann equation for
the Chern-Simons problem\cite{Ady,Kim} pointed out that in
Eq.~\ref{eq:frules}, $\frac{d \mu}{d n}$ is taken at fixed $\Delta B$.
This yields an $f_0$ (as well as an $f_1$) on the scale of ${\cal
  O}(m_b^{-1})$ in the limit $m_b \rightarrow 0$.  We will now show
how an ${\cal O}(m_b^{-1})$ value of $f_0$ is consistent with our
attachment of magnetization.  Fixing $\Delta B$ at zero means that
when we add a fermion to the system, the external field must also be
increased by a total of $\tilde \phi$ flux quanta.  Thus, the magnetic
field is linked to the density $n$ via $B = \tilde \phi n \Phi_0$.
The interaction energy between the magnetization ${\bf  M} = \mu_b n$
and the external field is given by $E = {\bf  M} \cdot {\bf  B} = \pi
\tilde \phi \hbar^2 n^2/m_b$.  Differentiating this with respect to
$n$ we obtain a magnetization contribution to the chemical potential
$\mu^{\mbox{\scriptsize{mag}}} = 2 \pi \tilde \phi \hbar^2 n/m_b=
\hbar \omega_c$ such that the magnetization contribution to $f_0$ is $
\tilde f_0 = d \mu^{\mbox{\scriptsize{mag}}}/d n = 2 \pi \tilde \phi
\hbar^2/m_b$ which is also the inverse compressibility of free
electrons of mass $m_b$ at constant $\Delta B$.  The coefficient $f_0$
is written $f_0 = \tilde f_0 + \delta f_0$ where $\tilde f_0$ is
${\cal O}(m_b^{-1})$ and $\delta f_0$ is on the smaller interaction
scale.  As mentioned in Ref.~\onlinecite{Ady}, in the limit $m_b
\rightarrow 0$, the requirement that the low energy spectrum is
independent of $m_b$ forces the other interaction coefficients ($f_l$
for $l \ne 0,1$) to be on the interaction scale\cite{LongerPaper}.

Since in the limit of $m_b \rightarrow 0$, $\tilde f_0$ and $f_1$ are
on the bare mass scale whereas all other coefficients $f_l$ (as well
as $\delta f_0$) are on the interaction scale, we will separate out
the contributions of these two coefficients by
writing\cite{LongerPaper}
\begin{equation}
  \label{eq:corrs}
  \tilde \Pi^{-1} = [\tilde \Pi^*]^{-1} + \tilde {\cal F}_0 + {\cal
    F}_1
\end{equation}
where $\tilde {\cal F}_0 = \mbox{diag}[\tilde f_0,0]$.  The function
$\tilde \Pi^*$ is to be calculated using a Landau-Boltzmann equation
representing quasiparticles with the same effective mass $m^*$ and
interaction coefficients $f_l$ except that $f_1$ is artificially set
to zero and the magnetic contribution $\tilde f_0$ is subtracted off
of $f_0$.  The separation of the coefficient $f_0$, equivalent to
taking $v(q) \rightarrow v(q) + f_0$ in Eq.~\ref{eq:KPi}, is achieved
by noting that $f_0$ corresponds to a short ranged density-density
interaction.  The separation of the nonzero $f_1$
coefficient\cite{Simonhalp,LongerPaper} is analogous to that described
in Eq.~\ref{eq:Pi*} and is derived in Ref.~\onlinecite{Simonhalp}.
Having made this separation, we expect that the response $\tilde
\Pi^*(q,\omega)$ is independent of $m_b$ in the limit $m_b \rightarrow
0$ and is well behaved for all values of $q/m_b$.

In Fermi liquid theory, the Landau-Boltzmann equation does not
correctly describe the Landau diamagnetic contribution to the
transverse static response.  Similarly, we suspect that here the
function $\tilde \Pi^*_{11}$ derived from the Landau-Boltzmann
equation lacks a term of the form $q^2 \chi$ where $\chi$ is some
appropriate Landau susceptibility which we expect to be on the scale
of the interaction strength. As usual, if we fix the ratio
$\omega/q$ to be nonzero, and take $q \rightarrow 0$, this diamagnetic
term becomes negligible.

Using the identities $\tilde {\cal F}_0 = M^{\dagger} \tilde {\cal
  F}_0 M$ and $U + \tilde {\cal F}_0 = M^\dagger{}^{-1} U M^{-1}$
(which holds in the limit $m_b \rightarrow 0$), we find that
$\mbox{M}^2$RPA is equivalent to approximating $\Pi^*$ by $K^{0*}$,
the response of a free Fermi gas of particles of mass $m^*$, and
calculating the response using Eqns.~\ref{eq:KPi}, \ref{eq:tildePi},
and \ref{eq:corrs}.

While the $\mbox{M}^2$RPA describes well the excitation spectrum of
fractionally quantized Hall states $\nu = \frac{p}{2mp+1}$ for large
$p$ at low wavevector $q$, it does not do so at high $q$.  In
particular, excitations at high $q$ are sensitive to the infra-red
divergence of the effective mass due to the gauge field
fluctuations\cite{Ady,Kim} which are neglected in $\mbox{M}^2$RPA.

To conclude, the transformation Eqns.~\ref{eq:KPi}, \ref{eq:Pi*},
\ref{eq:F1def}, \ref{eq:tildePi}, and \ref{eq:corrs} do not in
themselves involve any approximations, and may be considered simply as
a means of defining a new `irreducible' response function $\tilde
\Pi^*(q, \omega)$.  Our claim that $\tilde \Pi^*$ is well behaved in
the limit $m_b \rightarrow 0$ we believe to be an exact statement
(although we have not proved it rigorously) independent of the
approximation used to define the $\mbox{M}^2$RPA.  (For this to be
true it does not matter whether we take the moment ${\mu_{\mbox{\tiny
      M}}}$ in Eq.~\ref{eq:Mdef2} to be precisely $\mu_b$ or whether
we include a correction of order $\mu_b m_b/m^*$.)

The approximation we introduce, the $\mbox{M}^2$RPA, describes the
$\nu=\frac{1}{2m}$ state as a Fermi liquid of magnetized composite
fermions with a finite renormalized effective mass $m^*$, an $f_1$
parameter dictated by Galilean invariance and an $f_0$ parameter
originating from the interaction of the magnetization with the
magnetic field.  All remaining Fermi liquid parameters (which are
expected to be on the much smaller interaction scale) are neglected.
The $\mbox{M}^2$RPA predicts the same $K_{00}$ as the MRPA, but in
contrast it yields the correct behavior for $K_{01}, K_{10}$, and
$K_{11}$ in the limit $m_b \rightarrow 0$ for arbitrary small $q$.  In
the limit $q \rightarrow 0$, for fixed $m_b \ne 0$, the
$\mbox{M}^2$RPA and MRPA become identical for all components of
$K_{\mu\nu}$.

We thank P.~A.~Lee and D.~Orgad for helpful discussions.  This
research was supported by NSF Grants No.~DMR-94-16910 and
DMR-95-23361.

\end{multicols}
\end{document}